\documentclass[onecolumn]{aastex6}

\usepackage{natbib}
\usepackage{color}
\usepackage{mathtools}
\usepackage{hyperref} 
\hypersetup{colorlinks=true,citecolor=blue}

\def    \apjl  		{\rm {ApJL}}

\def	\cm		{\,{\rm {cm}}}

\def	\H		{{\rm H}}

\def	\erg	{\,{\rm {erg}}}

\def    \exp 	{\,{\rm {exp}}}

\def	\s		{\,{\rm s}}

\newcommand     \mum    {\,\mu{\rm m}}  

\def \bea {\begin{eqnarray}}
\def \ena {\end{eqnarray}}

\def	\cm		{\,{\rm {cm}}}

\def	\erg		{\,{\rm {erg}}}

\def    \exp 		{\,{\rm {exp}}}

\def	\K		{\,{\rm K}}

\def	\AU		{\,{\rm {au}}}

\def	\s		{\,{\rm s}}

\def	\H		{\rm H}

\def    \apjl  		{\rm {ApJL}}

\begin{document}

\title{Detectability of Thermal Emission from Sub-Relativistic Objects}

\author{Thiem Hoang}
\affil{Korea Astronomy and Space Science Institute, Daejeon 34055, Republic of Korea; \href{mailto:thiemhoang@kasi.re.kr}{thiemhoang@kasi.re.kr}}
\affil{Korea University of Science and Technology, Daejeon 34113, Republic of Korea}
\author{Abraham Loeb}
\affil{Astronomy Department, Harvard University, 60 Garden Street, Cambridge, MA, USA; \href{mailto:aloeb@cfa.harvard.edu}{aloeb@cfa.harvard.edu}}



\begin{abstract}
We calculate the surface temperature and the resulting brightness of sub-relativistic objects moving through the Solar system due to collisional heating by gas and radiative heating by solar radiation. The thermal emission from objects of size $\gtrsim 100$ m and speed of $\gtrsim 0.1c$, can be detected by the upcoming {\it James Webb Space Telescope} out to a distance of $\sim 100$ au. Future surveys could therefore set interesting limits on the abundance of fast-moving interstellar objects or spacecraft.
\end{abstract}

\section{Introduction}

Propulsion of lightweight spacecraft to sub-relativistic speeds that greatly exceed $30 ~\rm km\s^{-1}$ (or $10^{-4}c$) attainable by chemical rockets is a promising frontier for space exploration (\citealt{Turyshev:2020to}; \citealt{2020arXiv200703530F}). 
Recently, \cite{2017ApJ...834L..20C} proposed a novel method to measure the mass of planets via interferometry by an array of relativistic spacecraft, envisioned by the Breakthrough Starshot initiative\footnote{https://breakthroughinitiatives.org/Initiative/3}. \cite{2018AcAau.152..370P} suggested a precursor that will launch slower spacecraft at $v\sim 0.01c$ to explore the Solar system. \cite{Witten:2020tc} proposed to use a sub-relativistic spacecraft of speeds $v\gtrsim 0.001c$ to indirectly probe Planet 9 through its gravitational influence on the spacecraft trajectory. \cite{2020ApJ...895L..35H} calculated the turbulent noise associated with drag and magnetic forces from the interstellar medium (ISM) at these high speeds. 

As relativistic spacecraft moves through the solar system, it is heated by its interaction with gas particles as well as solar wind and radiation (\citealt{2017ApJ...837....5H}). Here, we study the detectability of the hot surface of a relativistic spacecraft from its long-wavelength thermal emission. In particular, we aim to derive the minimum size and speed of the spacecraft that could be detected with a high-sensitivity telescope.

The structure of the paper is as follows. In Section \ref{sec:model}, we describe the model of calculations, including spacecraft heating, cooling, and its thermal emission. In Section \ref{sec:results}, we present our numerical results of the surface temperature and emission flux, and discuss the minimum sizes of objects detectable by telescopes. In Section \ref{sec:discuss}, we discuss implications of our findings.



\section{Model}\label{sec:model}
Let us first evaluate the equilibrium temperature of the spacecraft due to collisional and radiative heating by absorption of interstellar starlight and solar radiation. Collisional heating is dominated by H and He which carry most of the gas mass. The rate of collisional heating to the spacecraft surface moving at a speed $v$ can be written as,
\bea
\frac{dE_{\rm coll}}{dt} = \frac{1.4n_{\H}m_{\H}v^{3}\pi R^{2}}{2}
\simeq 9.86\times 10^{4}\left(\frac{n_{\H}}{1\cm^{-3}}\right) \left(\frac{R}{1\cm}\right)^{2}\left(\frac{v}{0.1c}\right)^{3} \erg\s^{-1},\label{eq:dEdt_coll}
\ena
where $\pi R^{2}$ is the frontal surface area (for a radius $R$) of the spacecraft, $n_{\rm H}$ is the hydrogen density, and a factor of $1.4$ accounts for the $10\%$ abundance of He relative to H. 

Beyond the heliopause, at a distance $D>D_{\rm hp}=100\AU$, from the Sun, the hydrogen density is $n_{\rm H}=n_{\rm ISM}$ with $n_{\rm ISM}$ the ISM density. Inside the heliosphere, $n_{\H}=n_{\rm sw}+n_{\rm neu}$ with $n_{\rm neu}$ being the neutral hydrogen density and $n_{\rm sw}$ being the proton density of the solar wind. We take $n_{\rm neu}= 0.1\cm^{-3}$ (\citealt{2009SSRv..146..235F}). The mean proton density of the solar wind at a distance $D$ is described by $n_{\rm sw}=n_{0}(D/1\AU)^{2}\cm^{-3}$ with $n_{0}\approx 6\cm^{-3}$ being the density at $D=1\AU$ (\citealt{Venzmer:2018gy}). Note that heating by heavy elements and dust is negligible on average because it carries only a small fraction of the gas mass.

The heating by cosmic rays (CRs) of particle energy $E$ and flux $F_{\rm CR}$ is given by,
\bea
\frac{dE_{\rm CR}}{dt} = F_{\rm CR}E\pi R^{2}\simeq 0.02\left(\frac{E}{1\rm GeV}\right)\left(\frac{F_{\rm CR}}{1\cm^{2}\s^{-1}}\right)\left(\frac{R}{1\cm}\right)^{2}\erg\s^{-1}.\label{eq:dECR}
\ena
The CR's  measured near the Earth have $F_{\rm CR}\approx 1\cm^{2}\s^{-1}$ (\citealt{1985A&A...144..147L}) and $E=1$ GeV, yielding subdominance of CR heating relative to collisional heating for $v\gtrsim 10^{-3}$c.
 
We assume that the local radiation field has the same spectrum as the interstellar radiation field (ISRF) in the solar neighborhood \citep{1983A&A...128..212M}, with a total radiation energy density of $u_{\rm MMP}\approx 8.64\times 10^{-13} \erg\cm^{-3}$, where the local radiation field is expressed as a factor, $U$, times the energy density of the ISRF, $u_{\rm MMP}$. The heating rate of the frontal surface by the background radiation is given by,
\bea
\frac{dE_{\rm abs}}{dt}=\pi R^{2}cUu_{\rm MMP}\epsilon_{\star},\label{eq:dErad}
\ena
where $\epsilon_{\star}$ is the surface absorption efficiency averaged over the starlight radiation spectrum.  

The local radiation field includes the averaged ISRF and the solar radiation, with
\bea
U=1+\frac{L_{\star}}{4\pi D^{2}cu_{\rm MMP}}=1+0.5\times 10^{8}\left(\frac{L_{\star}}{L_{\odot}}\right)\left(\frac{1\AU}{D}\right)^{2},
\ena
implying an enhancement by the Sun within $D=10^{4}\AU$. 

The surface of the spacecraft at a temperature $T_{s}$ also emits thermal radiation, which results in radiative cooling at a rate:
\bea
\frac{dE_{\rm emiss}}{dt}=\int d\nu \pi R^{2}Q_{\rm abs, \nu}\pi B_{\nu}(T_{s})
=\pi R^{2}\epsilon_{T} \sigma T_{s}^{4},~~~~
\label{eq:dEcdt}
\ena
where,
\bea
\epsilon_{T}=\frac{\int d\nu Q_{\rm abs,\nu}B_{\nu}(T_{s})}{\int d\nu B_{\nu}(T_{s})}, \label{eq:Qabsavg}
\ena
is the Planck-spectrum averaged absorption efficiency, and $Q_{\rm abs}$ is the absorption efficiency of the spacecraft material.
 
The energy balance between radiative heating and cooling yields the surface equilibrium temperature,
\bea
T_{\rm s}=\left(\frac{cUu_{\rm MMP}+1.4 n_{\H}m_{\H}v^{3}/2+F_{\rm CR}E_{\rm CR}}{\sigma}\right)^{1/4}\left(\frac{\epsilon_{\star}}{\epsilon_{T}}\right)^{1/4}.\label{eq:Tsp}
\ena

For the ISM and $v\gtrsim 10^{-3}c$, collisional heating dominates. Equation (\ref{eq:Tsp}) yields the surface temperature,
\bea
T_{\rm s}\simeq \left(\frac{1.4 n_{\H}m_{\H}v^{3}}{2\sigma} \right)^{1/4}\left(\frac{\epsilon_{\star}}{4\epsilon_{T}}\right)^{1/4}\simeq 272.6\left(\frac{n_{\rm H}}{10\cm^{-3}}\right)^{1/4}\left(\frac{v}{0.1c}\right)^{3/4}\left(\frac{\epsilon_{\star}}{\epsilon_{T}}\right)^{1/4}\rm K,\label{eq:Tsp_coll}
\ena
implying that the emission spectrum peaks at a wavelength,
\bea
\lambda_{\rm max}=\frac{b}{T_{\rm sp}}\simeq 10.6\left(\frac{n_{\rm H}}{10\cm^{-3}}\right)^{-1/4}\left(\frac{v}{0.1c}\right)^{-3/4}\left(\frac{\epsilon_{\star}}{\epsilon_{T}}\right)^{-1/4}\mum,
\ena 
with $b$ is Wien's constant.

When the spacecraft enters the Solar system, radiative heating dominates. Equation (\ref{eq:Tsp}) yields the equilibrium temperature
\bea
T_{\rm s}\simeq 388.7\left(\frac{D}{1\AU}\right)^{-1/2}\left(\frac{\epsilon_{\star}}{\epsilon_{T}}\right)^{1/4}\K,\label{eq:Tsp_rad}
\ena
implying a peak wavelength
\bea
\lambda_{\rm max}=\frac{b}{T_{\rm s}}\simeq 7.45\left(\frac{D}{1\AU}\right)^{1/2}\left(\frac{\epsilon_{\star}}{\epsilon_{T}}\right)^{-1/4}\mum.
\ena 

Equations (\ref{eq:Tsp_coll}) and (\ref{eq:Tsp_rad}) imply that radiative heating by the Sun dominates over collisional heating at $D<10\AU$ for a speed $v=0.1c$. 

The bolometric radiation flux observed on Earth from the spacecraft is,
\bea
F=\frac{1}{4}\theta^{2}\sigma T_{\rm s}^4=\frac{1}{4}\left(\frac{R}{D}\right)^{2}\sigma T_{\rm s}^{4},
\ena
where $\theta=(R/D)$ is the angular size of the spacecraft on the sky, namely its radius, $R$, divided by its distance, $D$. For a temperature of $T_{s}=100\K$, the bolometric flux is,
\bea
F\simeq 6.3\times 10^{-24} \left(\frac{R}{1\rm m}\frac{100\AU}{D}\right)^{2}\left(\frac{T_{\rm s}}{100\rm K}\right)^{4}~\frac{\rm erg}{\cm^{2}\s}.
\ena

The spectral flux density is given by 
\bea
F_{\nu}=\frac{1}{4}\theta^{2}\pi B_{\nu}(T_{\rm s})=\frac{1}{4}\left(\frac{R}{D}\right)^{2}\frac{2\pi h\nu^{3}}{c^{2}}\frac{1}{\exp(h\nu/kT_{\rm s})-1}.\label{eq:Sflux}
\ena




\section{Results}\label{sec:results}
\begin{figure}
\includegraphics[width=0.5\textwidth]{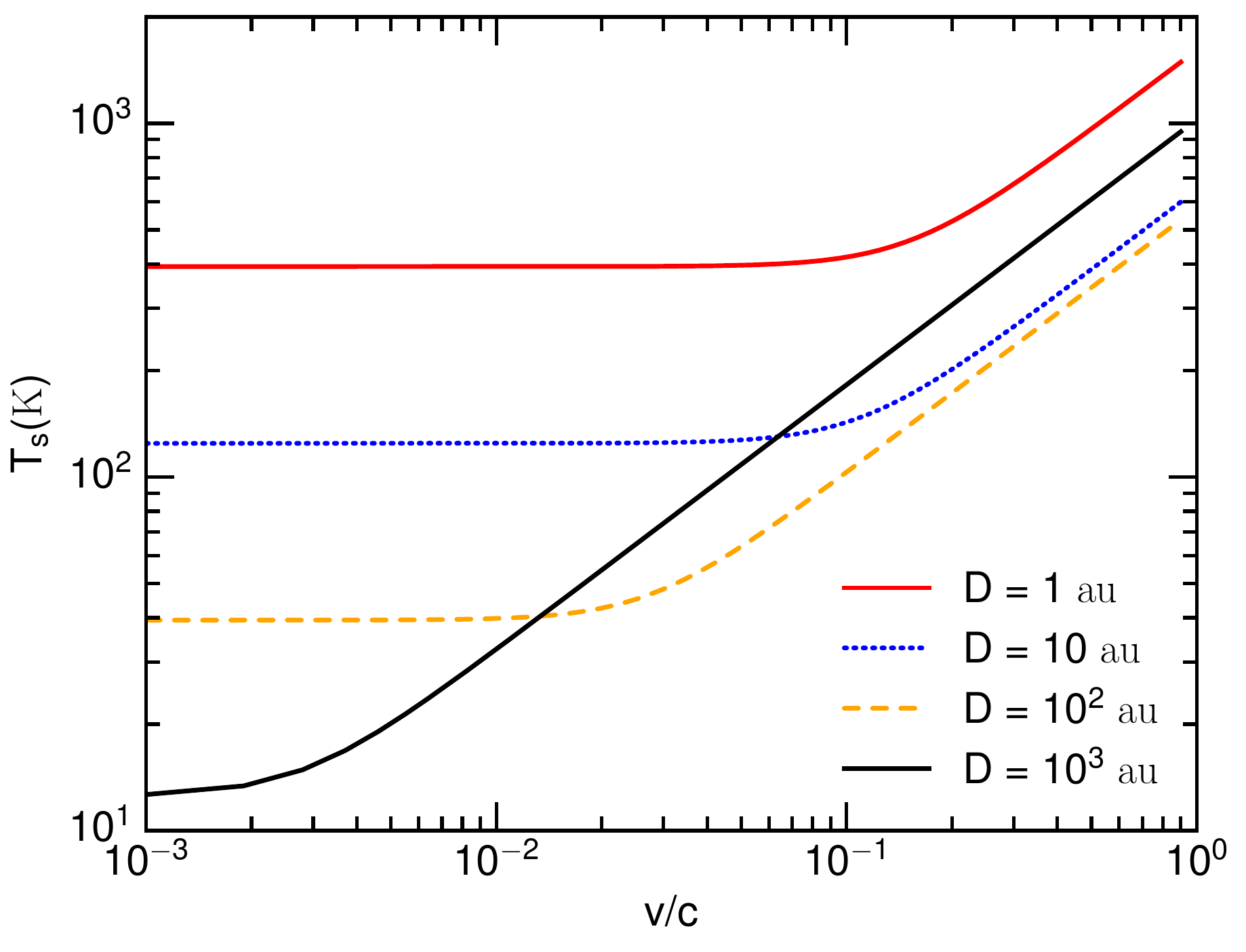}
\includegraphics[width=0.5\textwidth]{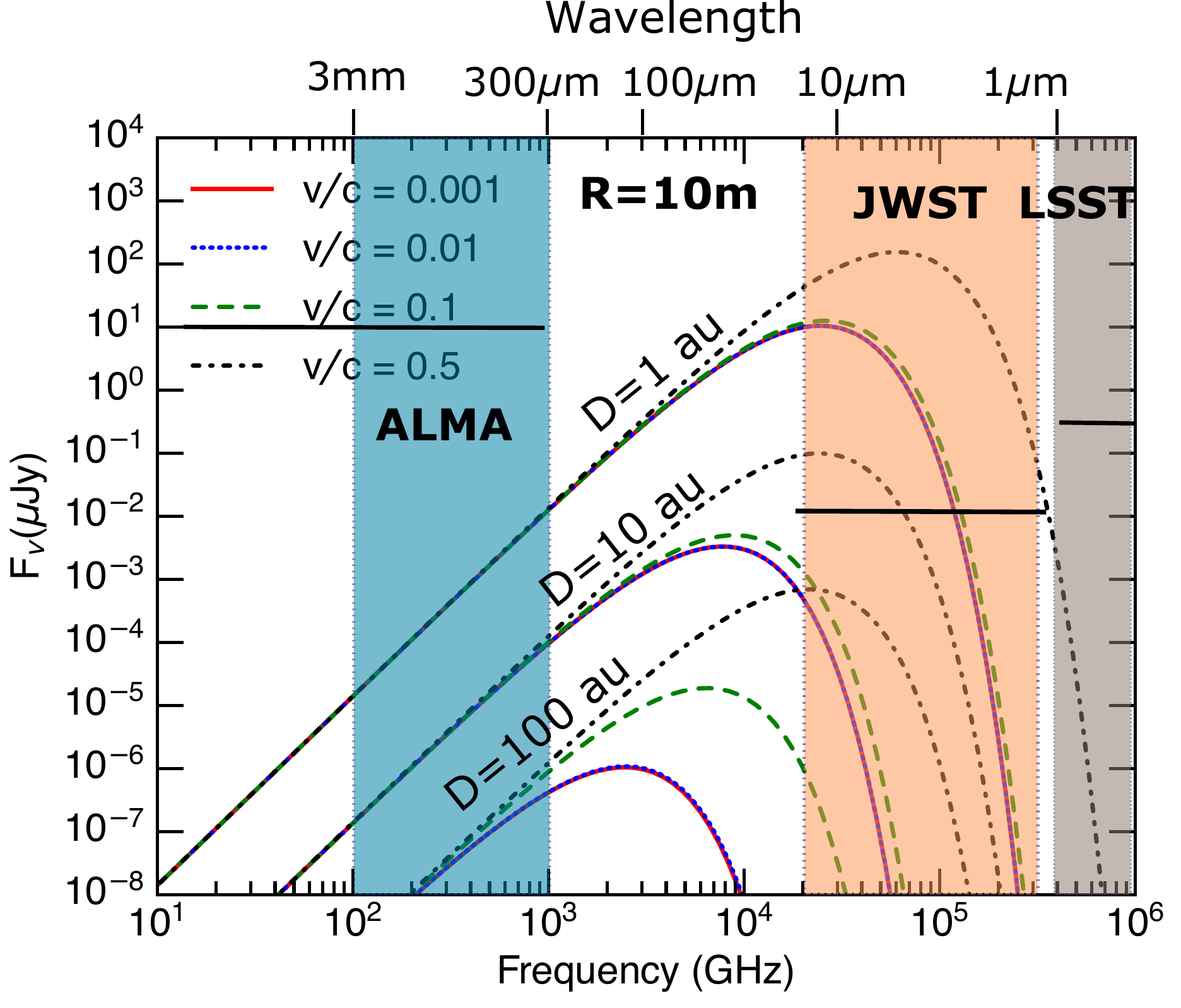}
\caption{Left panel: Surface temperature of the spacecraft as a function of its speed relative to the speed of light, $v/c$, at different heliocentric distances, $D$. Right panel: Spectral flux density for different spacecraft speeds, assuming the object radius $R=10$ m and $D=1,10, 100$ au. Horizontal black lines mark the sensitivity of various telescopes.}
\label{fig:Teq}
\end{figure}

The left panel of Figure \ref{fig:Teq} shows the surface temperature as a function of spacecraft speed in units of the speed of light, ($v/c$), for different heliocentric distances, and assuming $\epsilon_{\star}/\epsilon_{T}=1$. The temperature is dictated by radiative heating at low speeds. When the speed is sufficiently large, collisional heating becomes important and $T_{\rm sp}$ increases with $v/c$. For example, at $D=10$ au, the temperature increases with speed for $v\gtrsim 0.02c$. At $D=1000$ au, the temperature at $v>0.005c$ is higher than that for $D=10, 100$ au due to the stronger collisional heating.

The right panel of Figure \ref{fig:Teq} shows the spectral flux density from Equation (\ref{eq:Sflux}) as a function of frequency (wavelength) for different spacecraft speeds, $v/c$, assuming a spacecraft radius $R=10$ m. The frequency bands of several telescopes are shown, including ALMA (Atacama Large Millimeter/submillimeter Array), JWST (James Webb Space Telescope), and LSST (Legacy Survey of Space and Time). The sensitivity is $S=8.7~\mu$Jy at $\lambda\approx 870\mum$ for ALMA\footnote{https://almascience.eso.org/proposing/sensitivity-calculator}, $0.01~\mu$Jy at $\lambda\sim 0.6-28.3\mum$ with integration time of $10^{4}$ s for JWST\footnote{https://www.stsci.edu/files/live/sites/www/files/home/news/newsletters/documents/2013-volume030-issue02.pdf} (\citealt{Anonymous:vWxVySok}; \citealt{2015PASP..127..686G}). LSST can reach the limiting magnitude of $m=25$ with two exposure times of 15 s (\cite{2019ApJ...873..111I}), yielding a sensitivity of $S=3631\times 10^{-m/2.5}~\rm Jy\sim 0.36\mu$J for the $\sim 0.3-1\mum$ wavelengths.

We find that only when a 10 meter-size objects enter the solar system, it is feasible to detect them using JWST. At a distance of 100 au, JWST can detect an object of size $\sim 400$ m moving at $v\gtrsim 0.1c$. For travel through the Solar wind near the Earth's orbit, i.e., at a distance of $D\sim$ 1 au, we get detectability for objects larger than $50$ cm. At $D=10\AU$, objects larger than $100$ m can be detected by JWST.

\section{Discussion}\label{sec:discuss}
Most sky surveys dismiss objects that move too fast on the sky. A sub-relativistic nearby object would move at an angular speed of $(v/D)=2.48(v/10^{-3}c)(10\AU/D)~{\rm arcsec/min}$, and this would limit the integration time within the field of view of telescopes. Hence the sensitivity will be lower than for steady point sources, unless a special tracking program is applied.

The situation improves for objects closer to the Sun, since the solar wind density increases inversely with the square of the distance from the Sun. The planned size of the transmitter or sails of Starshot spacecraft is a few meters and the goal of Starshot is to probe the habitable zone around a star, similar to the region around the Earth's orbit. Thus, the mid- and far-infrared special regimes is optimal for detecting analogous spacecraft in our vicinity.


\acknowledgments 
This work is supported in part by a grant from the Breakthrough Prize Foundation.

\bibliography{ms.bbl}

\end{document}